\documentclass[referee]{mn2e}

\usepackage{graphicx}
\usepackage{amsmath}

\title[Gamma-rays from pulsar population in the Galactic Center]
{Gamma-rays from millisecond pulsar population within the central stellar cluster in the Galactic Center}
\author[W. Bednarek \& T. Sobczak]
{W. Bednarek \& T. Sobczak\\
Department of Astrophysics, The University of Lodz,
ul. Pomorska 149/153, 90-236 Lodz, Poland,\\
bednar@astro.phys.uni.lodz.pl, tmsobczak@uni.lodz.pl}
\begin{document}

\date{Accepted . Received ; in original form }

\pagerange{\pageref{firstpage}--\pageref{lastpage}} \pubyear{2013}

\maketitle

\label{firstpage}

\begin{abstract}
It was proposed that the central dense stellar cluster in the Galactic Center, containing the mass of $\sim$4 times larger than that of the central black hole, had been formed as a result of a merging of several massive globular clusters. These globular clusters are expected to provide a large number of millisecond pulsars (MSPs) within the central parsec of the Galactic Center. We propose that the GeV $\gamma$-ray emission observed from the Galactic Center is in fact a cumulative effect of the emission from several globular clusters
captured by the Galactic Center black hole. Moreover, the millisecond pulsars in globular clusters accelerate leptons in their wind zones to energies of the order of a few tens of TeV injecting them into the  dense infrared and optical radiation region present within the central parsec. We calculate the expected TeV $\gamma$-ray emission produced by these leptons by the Inverse Compton Scattering process in the soft radiation field. It is shown that this emission can be responsible for the multi-TeV $\gamma$-rays observed  by the Cherenkov telescopes from the Galactic Center for reasonable densities of the soft radiation, diffusion models for the propagation of leptons, their injection parameters (fluxes and spectral proprieties). If the energy conversion efficiency from the pulsars to the relativistic leptons is of the order of $10\%$, then about a thousand of MSPs have to be present in the central cluster in the Galactic Center.
\end{abstract}

\section{Introduction}

After the discovery of the supermassive black hole within the center of the Galaxy with the mass of the order of that expected in some active galaxies, it has been suspected that the Galactic Center (GC) should be the site of acceleration of particles to high energies. In fact, $\gamma$-ray emission was detected from the source 3EG J1746-2851 towards GC in the GeV energy range already by the EGRET telescope (Hartman et al.~1999, Mayer-Hasselwander et al.~1998). The existence of this source was  recently confirmed by the Fermi-LAT observations (Abdo et al.~2010a). 
Observations with various Cherenkov
telescopes discovered also the $\gamma$-ray emission extending up to several TeV (Tsuchiya et al.~2004, Kosack et al.~2004, Aharonian et al.~2004, Albert et al.~2006). Although GC varies strongly at lower energies, the GeV-TeV flux seems to be steady in the time period of several years.
The $\gamma$-ray emission from GC shows two distinct components since an extrapolation of the multi-GeV spectrum does not meet the level of the TeV emission (Chernyakova et al.~2011). The TeV spectrum 
is well described by a power law (differential spectral index $2.1\pm 0.04_{\rm stat}\pm 0.10_{\rm  syst}$) with an exponential cut-off at $15.7\pm 3.4_{\rm stat}\pm 2.5_{\rm syst}$ TeV (Aharonian et al.~2009), extending up to a few tens of TeV (Aharonian et al.~2004).  
The TeV source is coincident with the position of the supermassive black hole Sgr A$^\star$ within an error circle of $13$ arcsec. An upper limit on the source size of 1.3 arc min, corresponding to $\sim$3 pc, was derived by Acero et al.~(2010). 

Different models usually relate the production of $\gamma$-rays to 
the existence of the central supermassive black hole (Sgr A$^\star$) or a plerion discovered within the central few arcseconds (Wang et al.~2006). In the first class of models, Sgr A$^\star$ is responsible for the acceleration of leptons (e.g. Atoyan \& Dermer~2004, Hinton \& Aharonian~2007, Kusunose \& Takahara~2012), hadrons (e.g. Liu et al.~2006, Ballantyne et al.~2011, Chernyakova et al.~2011, Linden et al.~2012, Fatuzzo \& Melia~2012), or both (e.g. Yi-Qing et al.~2013).
In this work, we consider the hypothesis in which the central stellar cluster in the GC is responsible for the high energy processes observed from the direction of Sgr A$^\star$.
Following the proposition of Wang et al.~(2005, see also Abazajian~2011) 
we assume that the GeV $\gamma$-ray emission is due to the cumulative effect of several globular clusters captured by the central black hole. In fact, $\gamma$-ray emission with very similar spectral features has been recently detected by Fermi-LAT from globular clusters (Abdo et al.~2010b).
The pulsar population within globular clusters should also efficiently accelerate leptons to multi-TeV energies (see Bednarek \& Sitarek~2007). Therefore, we propose that the TeV $\gamma$-rays from the direction of GC are produced by leptons with multi-TeV energies which Inverse Compton up-scatter the infrared and optical background in the central stellar cluster around the supermassive black hole identified with the Sgr A$^\star$.

\section{Central stellar cluster in GC}

The central region of the Galaxy contains a stellar cluster with the mass of $\sim 1.5\times 10^7$ M$_\odot$ surrounding the central supermassive black hole with the mass $\sim 4\times 10^6$ M$_\odot$. 
The nature of this cluster is not clear.
It is argued that such cluster could appear as a result of a merger of several globular clusters (GlCs) due to dynamical friction (Antonini et al.~2012, Capuzzo-Dolcetta et al.~2013). The central cluster has a uniform core with the radius of $R_{\rm c} = 0.5$ pc and the stellar density profile outside the core described by the function $\rho\propto r^{-1.8}$ up to $\sim 30$ pc.
It is well known that globular clusters contain large populations of millisecond pulsars (MSPs). In fact, about half of observed MSPs were discovered within globular clusters (Harris 1996). It is estimated that some globular clusters may contain up to a few hundred MSPs. These expectations are consistent with the recent observations of the GeV $\gamma$-ray emission from several GCs by the Fermi-LAT telescope (Abdo et al. 2009, Abdo et al. 2010b, Kong et al.~2010, Tam et al.~2011).
If the above hypothesis is correct, then the central stellar cluster could contain up to a few tausend MSPs which might be responsible for the observed GeV emission from the Galactic Center (see the model by Wang et al.~2005).

We propose that MSPs in the GC can accelerate leptons to $\sim 100$ TeV in their wind regions. In fact, efficient acceleration of particles to energies of this order is observed in the case of the winds around several classical pulsars. The acceleration in the winds of MSPs is expected to occur similarly since the processes in the inner magnetospheres of both groups of pulsars are expected to be the same (similar spectral features of the pulsed GeV emission).
Note that a similar scenario has been recently considered in the models for the TeV $\gamma$-ray emission from GlCs (e.g. Bednarek \& Sitarek~2007, Venter et al.~2009, Cheng et al.~2010). 

In the next section, we constrain the model for acceleration of leptons in GC. Their diffusion process within the central parsec of the Galaxy and their interaction with the radiation field are considered. For the basic parameters determining such model we adopt those generally consistent with the observational constraints on the magnetic field strength in the central region of the GC ($B > 50\mu$G, Crocker et al.~2010) and on the GC infrared and optical radiation field (energy densities $\sim 10^4$ eV cm$^{-3}$, Mezger et al.~1996, Kusunose \& Takahara~2012). 
Leptons injected into such environment diffuse from the central cluster losing energy on the synchrotron and Inverse Compton (IC) radiation. 
Our calculations show that the TeV $\gamma$-ray emission observed from the GC up to a few tens of TeV can be explained by the IC emission from leptons, accelerated in the winds of MSP, which up-scatter the diffusive soft radiation field in the GC.

\section{Radiation from leptons}

In our model, MSPs are responsible for the acceleration of leptons in the central stellar cluster in GC. A MSP creates a strong pulsar winds with the  power defined by the pulsar basic parameters, 
\begin{eqnarray}
L_{\rm p}\approx 3.6\times 10^{34}B_8^2/P_3^4~~~{\rm erg~s^{-1}},
\label{eq1}
\end{eqnarray}
\noindent
where $B_{\rm pul} = 3\times 10^8B_8$ G is the pulsar surface magnetic field strength, and $P_{\rm pul} = 3P_3$ ms is the pulsar rotational period. The mean distance between MSPs within the central cluster in GC can be estimated as,
\begin{eqnarray}
D = 3\times 10^{17}R_1N_3^{-1/3}~~~{\rm cm},
\label{eq2}
\end{eqnarray}
\noindent
where $R_{\rm c} = 1R_1$ pc is the core radius of the central cluster, and $N = 10^3N_3$ is the number of MSPs within the central cluster. 
The MSPs move within the core of the central cluster with typical rotational velocities in the gravitational potential of the cluster of the order of,
\begin{eqnarray}
v_{\rm r} = \sqrt{2GM/R}\approx 300 (M_7/R_1)^{1/2}~~~{\rm km~s^{-1}},
\label{eq3}
\end{eqnarray}
\noindent
where $G$ is the gravitational constant, $M = 10^7M_7$ M$_\odot$ is the mass of the cluster within $R$.

The wind of MSP interacts with the cluster medium creating a bow shock
with the radius, $R_{\rm sh}$, that can be determined from the comparison of the kinetic energy density of the medium (relative to pulsar) and the energy density of the pulsar wind,
\begin{eqnarray}
L_{\rm p}/4\pi cR_{\rm sh}^2 = \rho v_{\rm r}^2/2,
\label{eq4}
\end{eqnarray}
\noindent
where $L_{\rm p} = 10^{34}L_{34}$ erg/s is the energy loss rate of the pulsar, $c$ is the velocity of light,  $R_{\rm sh}$ is the shock radius, $\rho = 10^3\rho_3$ cm$^{-3}$ is the density of matter in the central cluster, and $v = 300v_3$ km s$^{-1}$ is the velocity of the pulsar. The distance of the shock from the pulsar is estimated from Eq.~3 and 4,
\begin{eqnarray}
R_{\rm sh} = 1.9\times 10^{14} L_{34}^{1/2}/\rho_3^{1/2}v_{3}~~~{\rm cm},
\label{eq5}
\end{eqnarray}
\noindent
i.e. it is much smaller than the characteristic distance between MSPs in the cluster core.
We estimate the magnetic field strength at the shock from the pulsar site by scaling it from the pulsar surface. The dipole structure of the magnetic field is assumed within the light cylinder radius ($B(r)\propto r^{-3}$) and its toroidal structure in the MSP wind region ($B(r)\propto r^{-1}$). Then, the magnetic field strength at the shock is of the order of,
\begin{eqnarray}
B_{\rm sh} = \eta B_{\rm LC}R_{\rm LC}/R_{\rm sh}\approx 7.5\times 10^{-4} \eta_{-1}B_8\rho_3^{1/2}v_{3}/(L_{34}^{1/2}P_3^2)~~~{\rm G},
\label{eq6}
\end{eqnarray}
\noindent
where $B_{\rm LC} = B_{\rm p}(R_{\rm NS}/R_{\rm LC})^3$ is the magnetic field strength at the light cylinder, $R_{\rm NS} = 10^6$ cm is the neutron star radius, $\eta = 0.1\eta_{-1}$ is the magnetization parameter of the pulsar wind, and $R_{\rm LC} =cP/2\pi$ is the light cylinder radius.

We assume that leptons are accelerated at the shock reaching a power law spectrum up to the maximum energy defined by the escape from the shock region or by their radiation energy losses. After all, the Larmor radius of leptons accelerated at the pulsar wind shock,
\begin{eqnarray}
R_{\rm L} =  cP/eB\approx 3\times 10^{13}E_{\rm TeV}/B_{-4}~~~{\rm cm}
\label{eq7}
\end{eqnarray}
\noindent
(where $E = 1E_{\rm TeV}$ is the lepton energy in TeV),  
has to be smaller than the extent of the shock, $R_{\rm sh}$. This allows us to put the upper limit on the lepton energy of the order of,
\begin{eqnarray}
E_{\rm L}\approx 48 \eta_{-1}B_8/P^2_3~~~{\rm TeV}.
\label{eq8}
\end{eqnarray}
\noindent
Note that this limit depends only on the pulsar parameters. 

The energies of leptons can be also limitted by their synchrotron energy losses. Their maximum energies can be estimated by comparing the electron acceleration rate, $\dot{E}_{\rm acc} = \xi cE/R_{\rm L}\approx 10^8\xi_{-1}B_{-4}$ eV s$^{-1}$ (where $\xi = 0.1\xi_{-1}$ is the acceleration coefficient), with the synchrotron energy loss rate, $\dot{E}_{\rm syn} = (4/3)c\sigma_{\rm T}U_{\rm B}\gamma^2\approx 35 B_{-4}^2E_{\rm TeV}^2$ eV s$^{-1}$. This comparison gives
$E_{\rm syn}^{\rm max}\approx 1.7\times 10^3 (\xi_{-1}/B_{-4})^{1/2}~~~{\rm TeV}$.
The acceleration of leptons may be saturated also by their energy losses on the Inverse Compton process in the Klein-Nishina regime for the case of the infrared and optical photons. We roughly estimate lepton energy losses in the Klein-Nishina (KN) regime by applying the value of the transition from between the Thomson (T) and KN regime,
$\dot{E}_{\rm T/KN} = (4/3)c\sigma_{\rm T}U_{\rm inf}\gamma_{T/KN}^2\approx 6.7\times 10^3~~~{\rm eV s^{-1}}$, 
where $\gamma_{T/KN} = m_{\rm e}c^2/\varepsilon_{\rm inf}\sim 5\times 10^6$ is the Lorentz factor of leptons corresponding to the transition from the T to KN regimes.
The energy density of the infrared photons (with energies $\varepsilon_{\rm inf}\sim 0.1$ eV) is estimated as $U_{\rm inf} = 10^{4}$ eV cm$^{-3}$ (Mezger et al.~1996), consistently with the observations of Davidson~(1992). The optical background, which is less important here due to the more energetic photons, is estimated as $10^{(4-5)}$ eV s$^{-1}$ (Mezger et al.~1996).
Therefore, the IC energy losses in the KN regime can not 
additionally constrain the maximum energies of leptons accelerated to energies given by Eq.~8.
We conclude that for likely parameters of the acceleration scenario (e.g. $\xi_{-1} = 1$ for the relativistic pulsar wind and $B\sim 10^{-4}$ G) the
limit due to energy losses of leptons are clearly less restrictive than the above limit due to the Larmor radii of particles. Therefore, we conclude that electrons can reach energies of the order of $\sim 100$ TeV for typical parameters of the MSPs within the central cluster (e.g. rotational period of 2 ms and surface magnetic field of $3\times 10^8$ G, see e.g. Wang et al.~2005).

Leptons accelerated at the MSP wind shocks with the power law spectrum extending up to energies of the order of a few tens of TeV diffuse outside the cluster interacting mainly with the infrared and optical radiation field of the central cluster. 
We consider three prescriptions for the diffusion model of leptons in the central cluster. In the first model, we assume that the magnetic field in the region between the shocks is highly turbulent. Therefore, the Bohm diffusion prescription seems to be appropriate. The diffusion coefficient is then given by, 
\begin{eqnarray}
D_{\rm Bohm}\approx 3\times 10^{23} (E_{\rm TeV}/B_{-4})~~~{\rm cm^2 s^{-1}.}
\label{eq11}
\end{eqnarray}
\noindent
In the second diffusion model, we apply a prescription for the diffusion coefficient appropriate for the Kolmogorov spectrum of turbulence which is more characteristic for the medium in the galactic disk. In this case, the diffusion coefficient has the form,
\begin{eqnarray}
D_{\rm Kol}\approx  1.3\times 10^{27}(E_{\rm TeV}/B_{-4})^{1/3}~~~{\rm cm^2 s^{-1}.}
\label{eq12}
\end{eqnarray}
\noindent
The diffusion coefficient has been normalized in the way described in Bednarek \& Protheroe~(2002), i.e. to the case when the Larmor radius is equal to the extend of the size of the diffusion region assumed to be equal to 3 pc.
In the third diffusion model we assume the Kolmogorov prescription for the diffusion coefficient but with its value one order of magnitude larger. 

We calculate the IC $\gamma$-ray spectra produced by leptons in the model discussed above. For this purpose we apply the Monte Carlo code developed for the $\gamma$-ray production in GlCs (Bednarek \& Sitarek~2007). We follow the propagation of leptons in the magnetic and radiation field of the central stellar cluster up to the distance of 3 pc from the center assuming that density of radiation is uniform within the core with the radius of 0.5 pc and for larger distances drop according to $\propto r^{-1.8}$
(see Capuzzo-Dolcetta et al.~2013). These leptons are injected with the differential power law spectrum (spectral index 2.) and the cut-off at 50 TeV (see Eq.~8) in the core of the cluster. We take into account the synchrotron energy losses of leptons during their propagation. The dependence of the IC $\gamma$-ray spectra on the parameters of the model, i.e. on the average magnetic field strength (a), on the density of the infrared radiation field (b), on the total optical luminosity of the central cluster (c), and on the diffusion model (d), is investigated in Fig.~1. We note that the IC $\gamma$-ray spectra start to decline from a simple power law already below $\sim$10 TeV for the magnetic field strengths above $\sim 10^{-4}$ G (see Fig.~1a). Therefore, in order to explain the features of the observed TeV $\gamma$-ray emission from the Sgr A$^\star$, the magnetic field should be not very far from the upper limit derived by Crocker et al.~(2010). The effect of the density of infrared and optical photons on the calculated spectrum is rather small
(see Figs.~1b,c). This is due to the fact that leptons already cool efficiently within the 3 pc core region of the central cluster for the whole range of the considered energy densities of the radiation field. However, low values of the infrared (radiation) field (clearly below $10^4$ eV cm$^{-3}$) seem to be excluded in our model since the TeV spectra start to cut-off below $\sim 10$ TeV (see Fig.~1b). The $\gamma$-ray spectral shape depends rather weakly on the density of optical photons in the central cluster. Note however, that the maximum in the TeV spectrum shifts to larger energies for a stronger optical radiation field which is due to a more efficient scattering of the optical photons by leptons in the KN regime (see Fig.~1c). The $\gamma$-ray spectra depend also weakly on the applied diffusion model of leptons within the central region of Galaxy. Only for the largest considered values of the diffusion coefficient, in the Kolmogorov model, the TeV flux (mainly around $\sim$100 GeV) starts to drop due to the escape of leptons from the cluster core (see Fig.~1d). Therefore, we conclude that the expected TeV $\gamma$-ray emission is not strongly dependent on the considered diffusion model of leptons.

\section{$\gamma$-ray spectrum from GC}

\begin{figure*}
\vskip 4.truecm
\includegraphics{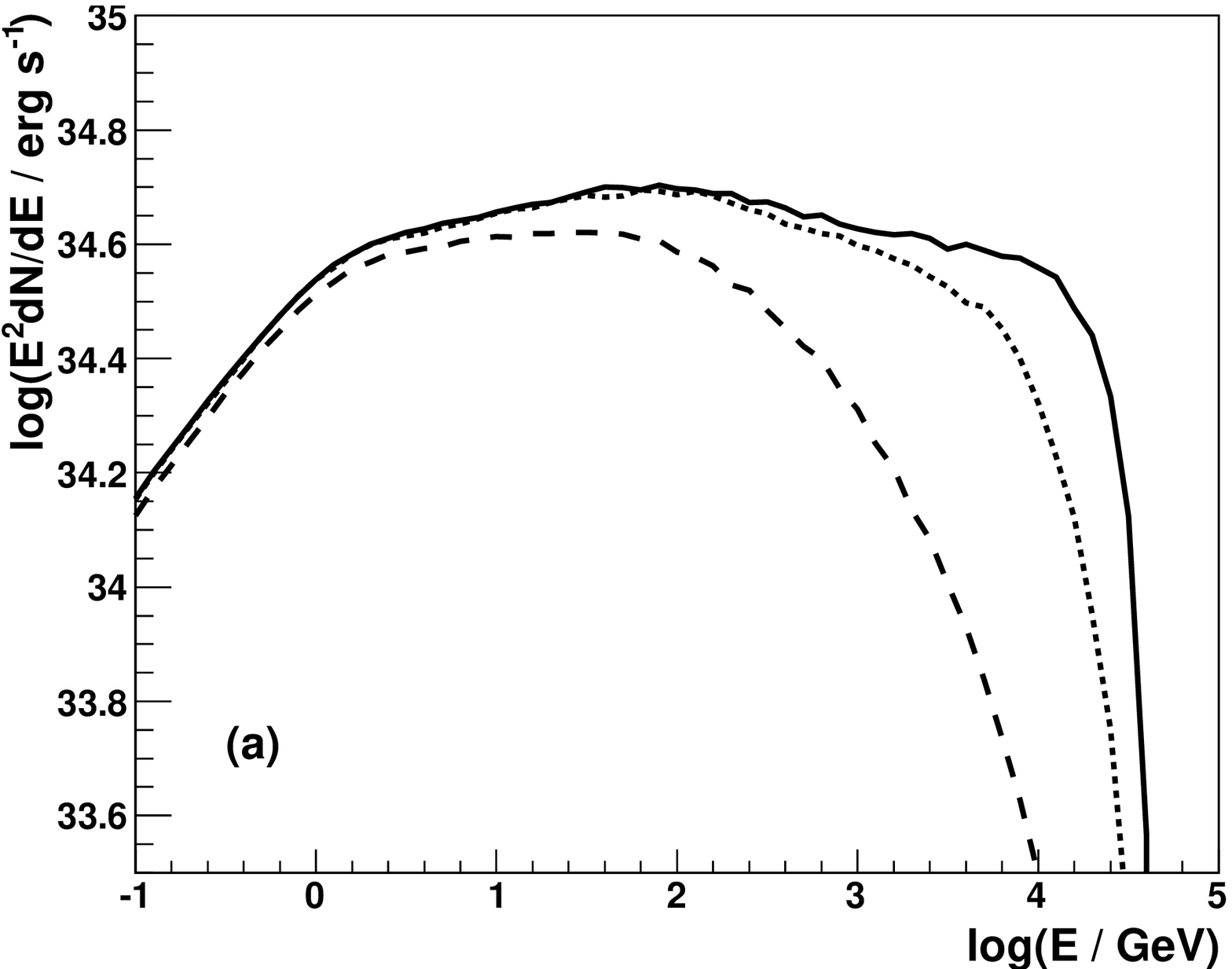}
\includegraphics{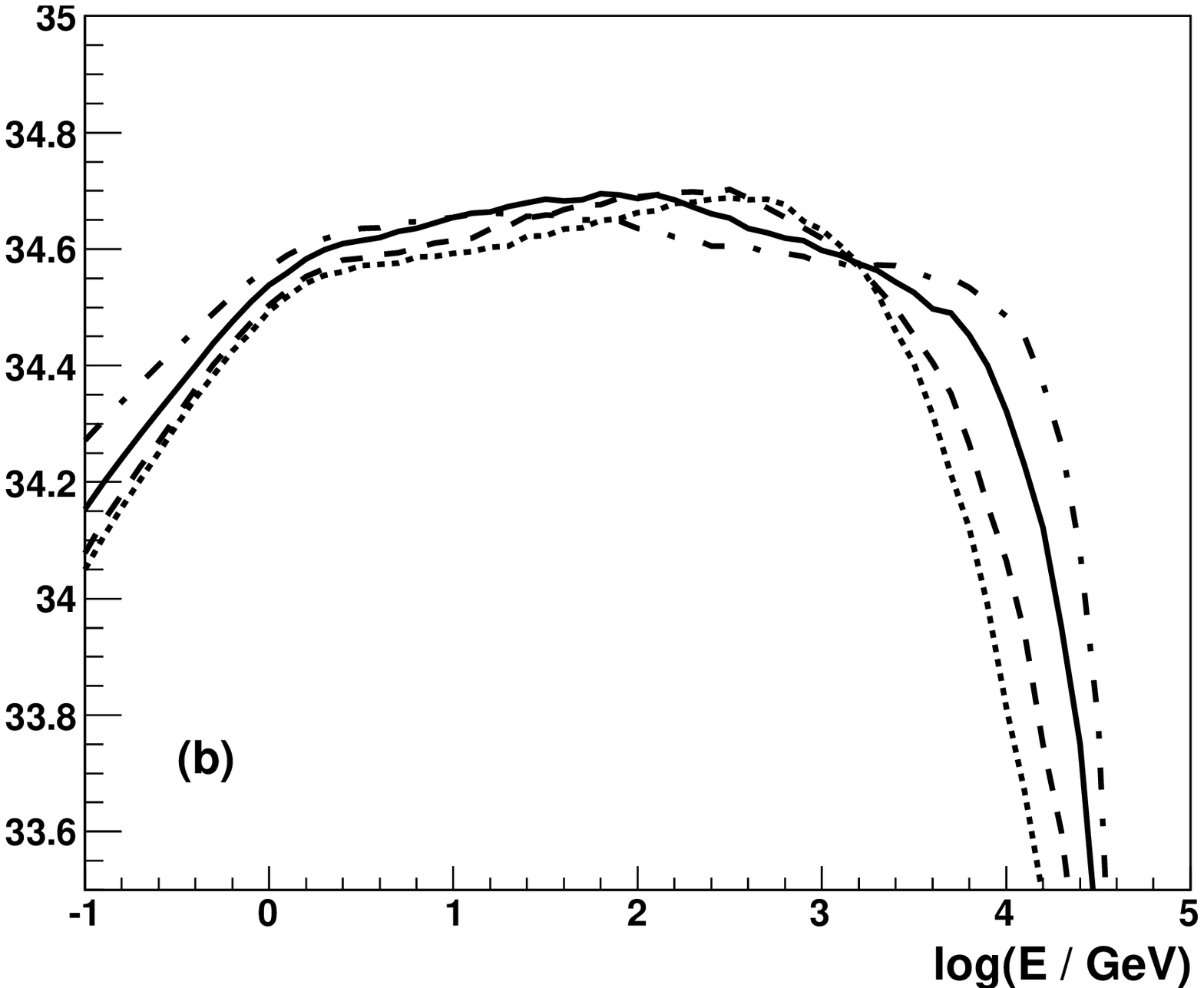}
\includegraphics{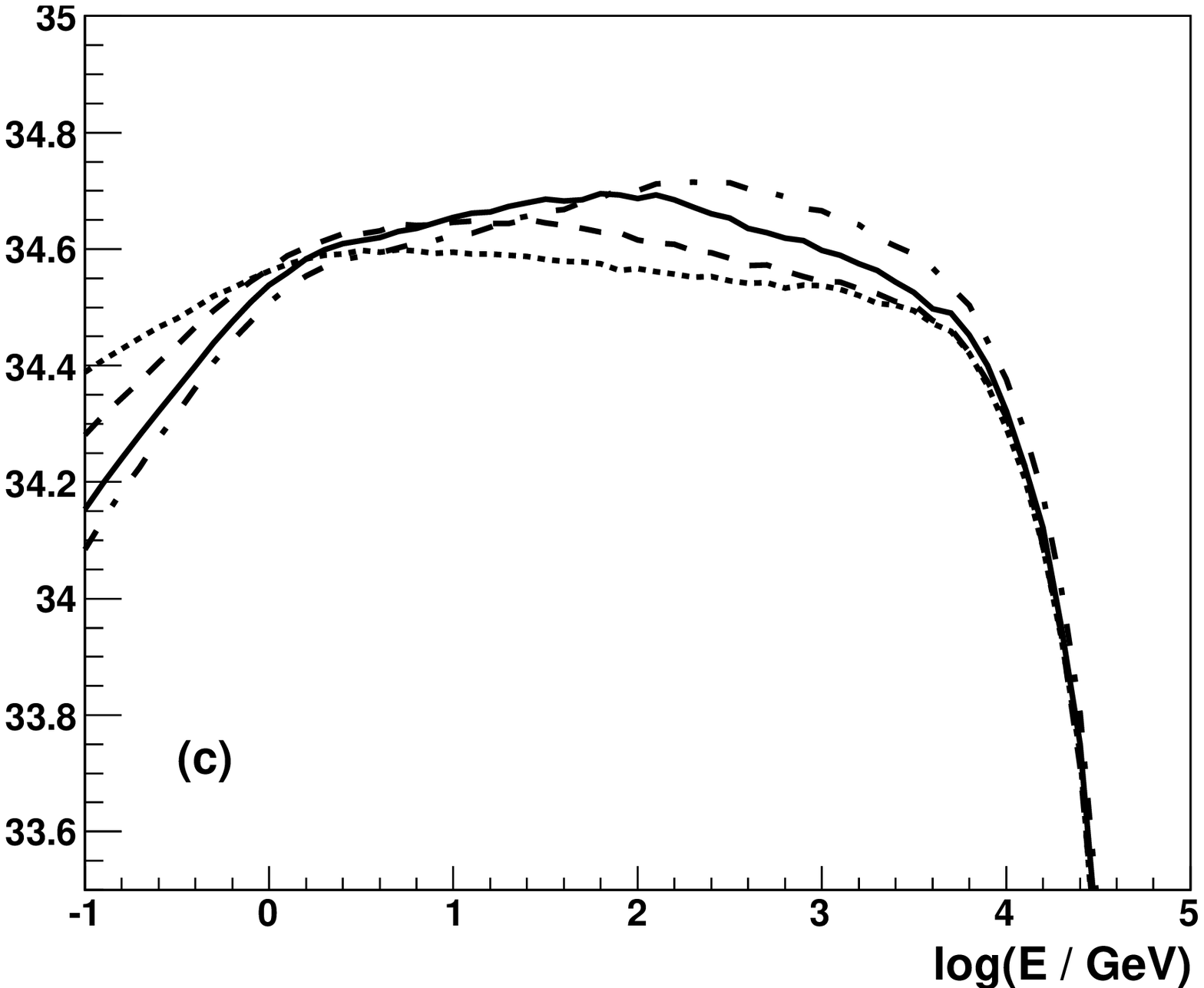}
\includegraphics{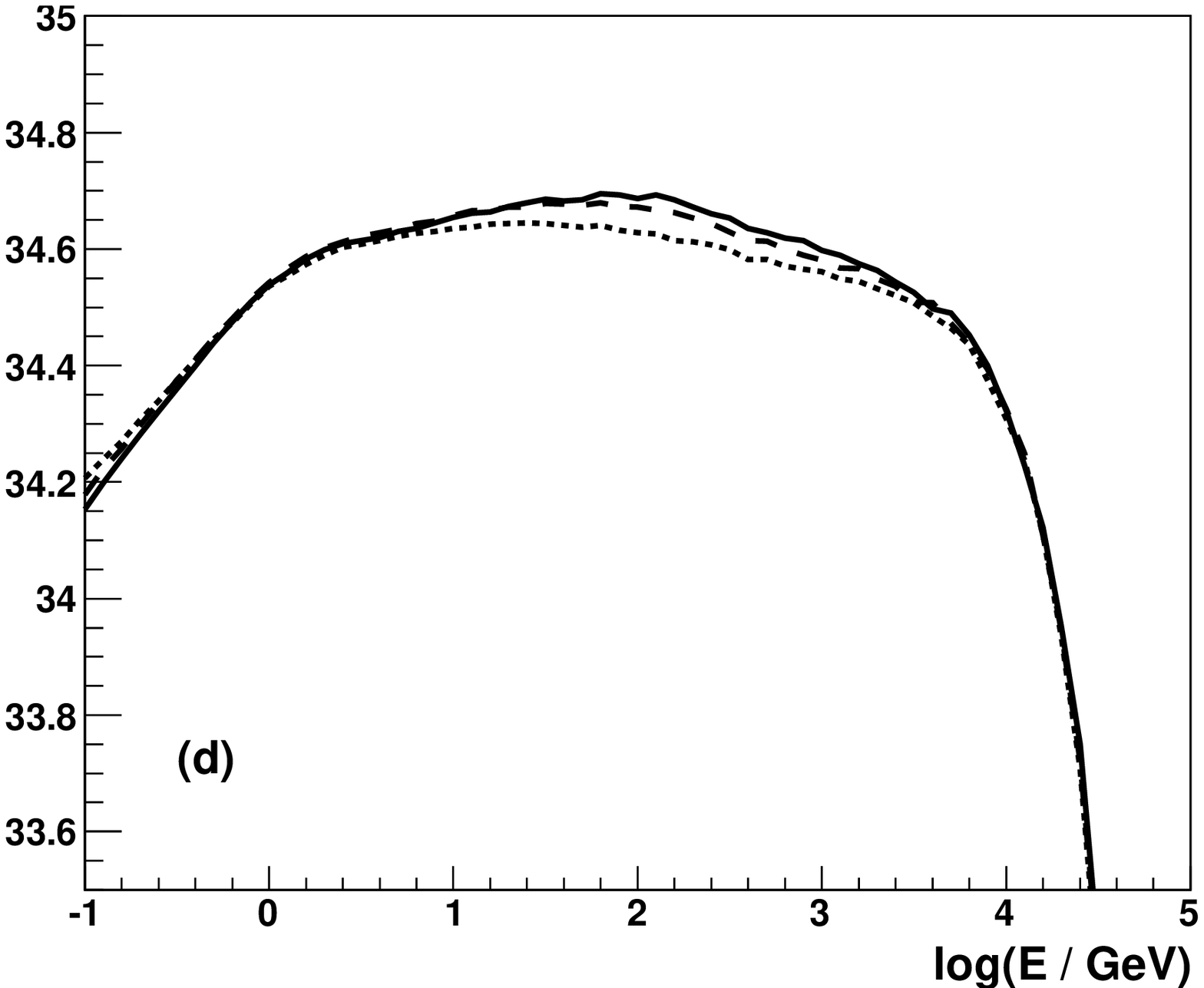}
\caption{Spectral energy distribution of the high energy IC $\gamma$-ray  emission from the Galactic Center region:
(a) dependence on the magnetic field: $B = 3\times 10^{-5}$ G (solid), $10^{-4}$ G (dotted), and $3\times 10^{-4}$ G (dashed), for other parameters fixed at  $L = 10^7$ L$_\odot$, $u_{\rm inf} = 10^4$ eV cm$^{-3}$, $R = 3$ pc and the Bohm diffusion model (Eq.~10); (b) dependence on the infrared radiation field with $T_{\rm inf} = 200$ K and  $u_{\rm inf} = 10^3$ (dotted), $3\times 10^3$ (dashed), $10^4$ (solid), and  $3\times 10^4$ eV cm$^{-3}$ (dot-dashed), and $B = 10^{-4}$ G, $L = 10^7$ L$_\odot$, $R = 3$ pc and the Bohm diffusion model; (c) dependence on the central cluster luminosity: $L = 10^6$ (dotted), $3\times 10^6$ (dashed) , $10^7$ (solid), and $3\times 10^7$ L$_\odot$ (dot-dashed) and $B = 10^{-4}$ G, $u_{\rm inf} = 10^4$ eV cm$^{-3}$, $R = 3$ pc; (d) dependence on the diffusion model of leptons within the central region: the Bohm diffusion  (solid), the Kolmogorov diffusion (Eq.~11, dashed), and with ten times larger diffusion constant (dotted). The other parameters for the last calculations are the following: $B = 10^{-4}$ G, $L = 10^7$ L$_\odot$, $u_{\rm inf} = 10^4$ eV cm$^{-3}$, $R = 3$ pc. In all calculations the spectrum of leptons is of the power law type with spectral index equal to 2. and the cut-off at 50 TeV.
}
\label{fig1}
\end{figure*}

We propose a two component origin of  the high energy $\gamma$-rays from the GC. 
The lower energy component, extending through the GeV energy range, is interpreted as due to the $\gamma$-ray emission from several globular clusters which create the central cluster as a result of the coalescence. This interpretation is supported by the recent discovery of GeV $\gamma$-rays from several GlCs (Abdo et al.~2010, Tam et al. 2011). In fact, the $\gamma$-ray spectra of some GlCs can be well described by a power law model with the exponential cut-off at a few GeV. However, the broken power law model also seems to work well for some GlCs especially for Ter 5, NGC 6652, M62 (see Fig.~2 in Abdo et al.~2010) or even a single power law model for NGC 6440 (Abdo et al.~2010). Moreover, Tam et al.~(2011) explicitly state that discovered by them $\gamma$-ray emission from GlCs can be  well described by a single power law model (see e.g. Fig. 5 for Liller 1 and Fig. 6 for NGC 6624 in that paper).
As an example, we show in Fig.~2 the shape of the $\gamma$-ray spectra from two GlCs, Ter 5  and M62 (see solid lines marked by "I" and "II" GC MSP), observed by Fermi-LAT (Abdo et al.~2010b), after renormalization to the $\gamma$-ray flux observed from the GC. 
The observed level of GeV emission from the GC (e.g. Chernyakova et al.~2011) can be understood in terms of such interpretation provided that MSPs from several GlC (with typical GeV $\gamma$-ray luminosity of a few $10^{35}$ erg s$^{-1}$, e.g. Ter 5, Lilier 1, or M62) are accumulated in the central stellar cluster. 

The second, higher energy component, in the $\gamma$-ray spectrum observed from the GC by Cherenkov telescopes is interpreted as due to the IC scattering of soft infrared and optical background radiation by relativistic leptons accelerated by this same population of MSPs. In Sect.~3, we argue that leptons can be accelerated to energies of the order of a hundred TeV (see Eq.~8). As an example, we compare the IC $\gamma$-ray spectrum produced by these leptons with a power law spectrum, with the spectral index equal to 2 extending up to 50 TeV with the HESS TeV measurements. The observed level of TeV $\gamma$-ray emission can be understood in terms of this interpretation provided that the product of the energy conversion efficiency from the MSPs to relativistic leptons, 
$\chi$, times the number of MSPs, $N_{\rm MSP}$, is of the order of 
$\chi\cdot N_{\rm MSP}\sim 110$, if the typical parameters of the MSPs are
$B_{\rm MSP} = 3\times 10^8$ G and $P_{\rm MSP} = 3$ ms. The value of $\chi$ is expected to be of the order of $10\%$ in the case of nebulae around classical  pulsars (e.g. the Crab Nebula). If the acceleration of leptons in winds of MSPs occurs similarly, then about a thousand of MSPs should be present in the central stellar cluster around Sgr A$^{\star}$.

Our Monte Carlo calculations of the TeV $\gamma$-ray spectra produced by leptons take also into account their energy losses in a relatively strong magnetic field in the central parsec around Sgr A$^\star$. 
The characteristic energies of the synchrotron photons produced by these leptons can be estimated from,
$\varepsilon_{\rm syn}\approx m_{\rm e}c^2(B/B_{\rm cr})\gamma_{\rm e}^2\approx 4.6B_{-4}E_{\rm TeV}^2$ eV. For leptons with the maximum energies considered in this model ($E_{\rm e} = 50$ TeV) and the magnetic field of the order of $B = 10^{-4}$ G, the synchrotron emission extends up to the X-ray range. However, only a relatively small part of the lepton energy goes into the synchrotron radiation since this process starts to dominate over their energy losses on IC process (in the KN regime) for energies above $\sim 14/B_{-4}$ TeV (estimated from energy loss rates given above). For leptons which scatter off soft photons in the Thomson regime the energy losses on IC dominate over those on synchrotron process by a factor of $\sim (30-300)/B_{-4}^2$. 
We find that the synchrotron X-ray emission from leptons accelerated by pulsars is consistent with the low state of the X-ray emission observed from central parsec around Sgr A$^\star$ by Chandra in the energy range 2-10 keV (Baganoff et al.~2003), provided that the average magnetic field strength in this region is not significantly larger than $\sim 100 \mu$G.

\begin{figure}
\vskip 5.truecm
\includegraphics{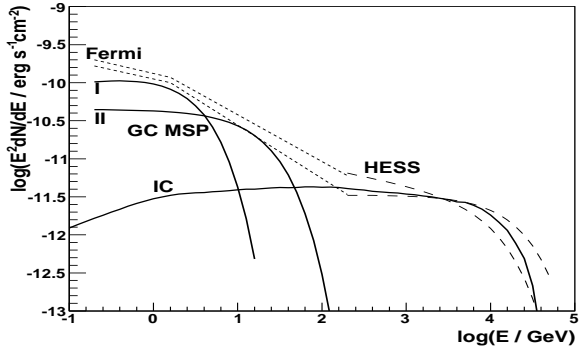}
\caption{Spectral energy distribution of $\gamma$-rays from the Galactic Center region measured by Fermi-LAT (dotted lines, Chernyakova et al. 2011) and the HESS Collaboration (dashed curves, Aharonian et al.~2009). The GeV $\gamma$-rays are interpreted as the contribution from the population of the millisecond pulsars present in the central region of Galaxy as a result of a merger of several globular clusters. As illustration, we show $\gamma$-ray spectra of two GlCs: Ter 5, solid curves marked by I, and M 62, marked by II (see Abdo et al.~2010), after renormalization to the GeV flux from the GC. The TeV $\gamma$-rays are produced by  IC of leptons accelerated in the MSP wind regions. The calculations have been done for: 
$B = 10^{-4}$ G, $L = 10^7$ L$_\odot$, $u_{\rm inf} = 10^4$ eV cm$^{-3}$, $R = 3$ pc, and the Bohm diffusion model with Eq.~10.
}
\label{fig2}
\end{figure}

In the model discussed above we consider also the extend of the TeV $\gamma$-ray source. It should be of the order of the diffusion distance of the largest energy electrons injected into the central cluster. This diffusion distance can be estimated from $R_{\rm dif} = \sqrt{2D\tau_{\rm IC}^{\rm T/KN}}$, where $D$ is the diffusion coefficient given by Eq.~(10) or Eq.~(11), and $\tau_{\rm IC}^{\rm T/KN} = E_{\rm e}/\dot{E}_{\rm T/KN}$ is the energy loss time scale of electrons in the KN regime, and the energy loss rate in the Klein-Nishina regime, $\dot{E}_{\rm T/KN}$, is given above. The largest energy losses of electrons are expected due to their scattering of the infrared radiation. For the parameters given above (i.e. $\varepsilon_{\rm inf}\sim 0.1$ eV and $U_{\rm inf} = 10^{4}$ eV cm$^{-3}$), the diffusion distance is estimated on
$R_{\rm dif}^{\rm Bohm}\approx  9.5\times 10^{15} E/B_{-4}^{1/2}$ cm in the case of Bohm diffusion and $R_{\rm dif}^{\rm Kol}\approx  6.2\times 10^{17} E^{2/3}/B_{-4}^{1/6}$ cm in the case of Kolmogorov diffusion. 
For the electrons with energies of 50 TeV, these diffusion distances are equal to $\sim 4.7\times 10^{17}$ cm and $\sim  8.5\times 10^{18}$ cm, respectively for the Bohm and Kolmogorov diffusion coefficients.
Note that these diffusion distances are smaller than (or comparable to)  the upper limit on the extend of the TeV $\gamma$-ray source at the Galactic Center estimated on 2.9 pc by the HESS Collaboration (Acero et al.~2010). We conclude that the extend of the IC source expected in our model is consistent with the observations of the TeV $\gamma$-ray source.

\section{Conclusion}

We conclude that the GeV $\gamma$-ray emission from the central parsec around Sgr A$^\star$  can be produced in the central cluster which is due to the capture of several globular clusters by the central black hole
(Antonini et al.~2012). We explain the steady TeV $\gamma$-ray emission from this region as being produced by relativistic leptons, accelerated in the MSP wind regions,
which up-scatter dense infrared and optical background radiation.
We estimate that in order to explain the observed level of the $\gamma$-ray emission from GC the product of the conversion of the rotational energy lost by the MSPs to relativistic leptons, $\chi$, and the number of MSPs in the central cluster, $N_{\rm MSP}$, has to be of the order of $\sim110$, provided that a typical pulsar magnetic field is $3\times 10^8$ G and the rotational period is $3$ ms (see Wang et al.~2005). The value of $\chi\sim 10\%$ (as observed in the case of e.g. the Crab Nebula) postulates the existence of about $10^3$ MSPs in the central cluster around Sgr A$^\star$.  We show that the extend of the TeV $\gamma$-ray source expected in our model is consistent with the upper limit on the source size derived by the HESS Collaboration (Acero et al. 2010). We perform the calculations of the TeV $\gamma$-ray spectrum by modifying
the Monte Carlo code developed for the TeV $\gamma$-ray emission from GlCs
(Bednarek \& Sitarek~2007). This version of the code takes into account the  synchrotron energy losses of leptons in the strong magnetic field within the central parsec. We conclude that the level of synchrotron emission from these leptons is consistent with the level of the low state X-ray emission observed by Chandra from the central parsec. 

Our model excludes any significant variability of the GeV-TeV $\gamma$ ray spectrum from the central region around Sgr A$^\star$ below the presently observed emission level. A higher $\gamma$-ray emission from this region could in principle be produced in the vicinity of the central black hole (due to a non-stationary accretion) or by one of variable point source in the GC (e.g. accreating neutron star or solar mass black hole). However, an observation of GeV-TeV $\gamma$-ray emission on the level clearly below the presently observed one would be inconsistent with our model.

\section*{Acknowledgments}
This work is supported by the grant through the Polish Narodowe Centrum Nauki No. 2011/01/B/ST9/00411.

%%%%%%%%%%%%%%%%%%%%%%%%%%%%%%%%%%%%%%%%%%%%

\label{lastpage}
\end{document}